# The empirical aspects of Eliashberg formalism for the superconducting mechanism in Iron-based superconductors

F. Shahbaz Tehrani and V. Daadmehr*

*Magnet & Superconducting Research Lab., Faculty of Physics & Chemistry, Alzahra University, Tehran 19938, Iran*

*Corresponding author:

Tel: (+98 21) 85692640 / (+98) 912608 9714

Fax: (+98 21) 88047861

E-mail: daadmehr@alzahra.ac.ir

URL: http:// www.alzahra.ac.ir/daadmehr/

First author:

E-mail: tehrani66@gmail.com

**Abstract**

We have investigated experimentally how properties of $NdFeAsO_{0.8}F_{0.2}$ superconductor affected due to the substitution of the $Ca^{2+}/Nd^{3+}$ doping. Based on the XRD data refinement, various structural parameters such as lattice parameters, bond angles, bond length, and etc. were studied. We have determined the upper limit of the calcium solubility in the $NdFeAsO_{0.8}F_{0.2}$ phase and it is restricted to $x \leq 0.05$. Also, we have found that the lattice parameters and the cell volume decreased by increasing the calcium content. According to the XRD data analysis, we have argued that these reductions are due to the variations in the bond lengths and the bond angles of (O/F)-Nd-(O/F) and As-Fe-As i.e. "α, β" upon increasing the calcium dopant. So, we have expected that the superconducting transition temperature ($T_C$) will be sensitive to the calcium doping values. Experimentally, the $T_C$ of our samples was reduced from 53 K (for x = 0) to 48 K (for x = 0.01) and 27 K (for x=0.025) and disappeared for our other sample. Then we have studied the dependence of $T_C$ and bond angles, bond length, the pnictogen height, and the lattice parameter to examine the available theories from an empirical point of view. The consistency of our experimental results and the theoretical reports based on the spin- and the orbital- fluctuation theories shows that these models play an important role in the pairing mechanism of the iron-based superconductors.

# I. Introduction

The discovery of the iron-based superconductors (FeSCs) continues more than a decade and these superconductors have attracted the attention of researchers due to the existence of their superconductivity and magnetic properties, simultaneously [1-6]. The FeSCs are one of the types of high-temperature superconductors and have a layered structure like cuprate superconductors[7-9]. Among of FeSCs, the 1111-type with a common formula of $ReFeAsO_{1-x}F_x$ (Re represents rare-earth element atoms) have the highest superconductivity transition temperature ($T_C$) and the FeAs layers act as the superconductive planes and the Re(O/F) layers are the charge reservoirs[10-12].

Doping at the Re(O/F) layers by substitution of the electron or hole dopants changes the $T_C$ of FeSCs and the effect of these substitutions is discovered on their electrical and magnetic properties[13-15]. The structural, electrical and magnetic properties of the calcium doped samples were studied in the 122- and 112-types of FeSCs[16-20]. However, the study and research about the calcium doping effects on the 1111-type are open such as the solubility limit, the role of the band length and band angles on the structural and superconductivity properties, the variations of the $T_C$ and etc.

The $T_C$ of the $NdFeAsO_{1-y}F_y$ compounds is higher than the other compounds of the 1111-type (the maximum $T_C$ has been reported for y=0.2 (Ref.21)). So, there are a lot of researches about the substitution effect of various ions on their structural and electrical properties[22-27]. But the calcium substitution effect on the structural and electrical properties of the $NdFeAsO_{0.8}F_{0.2}$ superconductor is not studied so far. So, we choose it for our study.

The FeAs4 tetrahedrons play an important role in the crystal structure of the 1111–type of FeSCs. T. Nomura et al. [28] investigated the distortion effect of the FeAs4 tetrahedron from its regular value ($\alpha$=109.47°) on the $T_C$ for the $AeFe_{1-x}Co_xAsF$ (Ae= Ca and Sr) compounds. C.H. Lee et al. [29, 30] have described that the $T_C$ of the $LnFeAsO_{1-y}$ (Ln=lanthanide) samples reduced when the bond angle moved away from the regular tetrahedron. Theoretically, this point has been offered by T. Saito et al. [31] using the orbital fluctuation theory in the iron pnictide superconductors. They have described that the electron-phonon coupling is the strongest when $\alpha$ tends to the regular value. Also, H. Usui et al. [32, 33] have attained the same result based on the spin-fluctuation theory and the fluctuation exchange (FLEX) method. Likewise, they have studied the dependence of the $T_C$ and the bond length of Fe-As. Moreover based on the spin- fluctuation theory and Eliashberg Equation, K. Kuroki et al. [34-36] have described the effects of “pnictogen height” above the Fe plane and the lattice parameter on the superconductivity. In addition, recently, many theoretical investigations have been carried out on the study of the pairing formalism in the FeSCs [37-39]. But, a conceptual and comprehensive comparison of the theoretical and experimental results about the effect of bond angles, bond length, the pnictogen height, and the lattice parameter on the $T_C$ has not been studied explicitly.

Given this background, in this work, we study the effects of $Ca^{2+}/Nd^{3+}$ substitution on polycrystalline $Nd_{1-x}Ca_xFeAsO_{0.8}F_{0.2}$ with $0 \leq x \leq 0.1$. We compare our experimental data and the results of the mentioned theories i.e. the spin- and orbital-fluctuation pairings. Specifically, we are going to analyze the structural properties of the doped samples via X-ray diffraction (XRD) by using the MAUD software. First, we try experimentally to determine the solubility of the calcium ions in the $NdFeAsO_{0.8}F_{0.2}$ superconductor. Second, we argue about the structural parameter such as bond angles, bond length, the pnictogen height, and microstrain comprehensively. Third, we measure the superconductivity properties and investigate the dependence of the $T_C$ and bond angles, bond length, the pnictogen height, and the lattice parameter. We hope that the correlation between our empirical results and the aforesaid theories will lead to an appropriate approach to a better understanding of the pairing mechanism in the FeSCs.

# II. Experimental

The polycrystalline samples with the nominal compositions of $Nd_{1-x}Ca_xFeAsO_{0.8}F_{0.2}$ with x=0.0, 0.01, 0.025, 0.05, and 0.1 were synthesized by one-step solid state reaction method as described in Ref. 21. The $NdFeAsO_{0.8}F_{0.2}$, $Nd_{0.99}Ca_{0.01}FeAsO_{0.8}F_{0.2}$, $Nd_{0.975}Ca_{0.025}FeAsO_{0.8}F_{0.2}$, $Nd_{0.95}Ca_{0.05}FeAsO_{0.8}F_{0.2}$ and $Nd_{0.9}Ca_{0.1}FeAsO_{0.8}F_{0.2}$ samples are labeled as Nd-1111, Nd-Ca0.01, Nd-Ca0.025, Nd-Ca0.05, and Nd-Ca0.1, respectively.

The X-ray diffraction patterns (XRD) of the synthesized samples were performed using a PANalytical® PW3050/60 X-ray diffractometer with Cu Kα radiation ($\lambda$= 1.54056 Å) operated at 40 kV and 40 mA with a step size of 0.026°. The refinement method of Rietveld was applied with the “Material Analysis Using Diffraction” (MAUD) software (v.2.8). A four probe technique was used for electrical transport measurements. The 20K Closed Cycle Cryostat (QCS101), ZSP Cryogenics Technology, was applied for superconductivity measurements. The applied DC current (Lake Shore-120) was 10 mA and the voltage was measured with microvolt accuracy. A Lake Shore-325 temperature controller was used for measuring the temperature.

# III. Results and discussion

### A. Structural study

The XRD patterns of the synthesized Nd-1111, Nd-Ca0.01, Nd-Ca0.025 and Nd-Ca0.05 samples have been shown in Fig. 1. The presence of the specified planes in the XRD patterns approves the formation of the tetragonal structure with the P4/nmm:2 space group in the all samples. Furthermore, the shift of the XRD-peaks to the major angle with the increase in the calcium content may be attributed to the reduction of the lattice parameter (the XRD-peaks (102) are shown in Fig.2). As shown in Fig. 1, there are two impurity phases of FeAs and NdOF in the Nd-1111 sample. In addition to the mentioned phases, we observe the $Nd_2O_3$ phase in the Nd-Ca0.01, Nd-Ca0.025, and Nd-Ca0.05 samples. These additional phases usually exist in the polycrystalline FeSCs that were reported in some previous studies[20-23, 40-43].

The XRD data of the synthesized samples have been refined by using the MAUD software with Rietveld's method based on its tetragonal structure with space group p4/nmm:2. The refinement results of our samples are presented in Table I. The agreement of the theoretical- and experimental-refinement results for the $Ca^{2+}$ and $Nd^{3+}$ ions in the doped samples ($x \leq 0.05$), illustrate the complete substitution of the calcium ions in the neodymium sites. The percent volume (Vol. %) of the phases for our synthesized samples are listed in Table II and has shown that the impurity phases have small amounts in all samples.

Figure 3 shows the XRD pattern of the synthesized Nd-Ca0.1 sample. The phases of $FeAs_2$, FeAs, NdOF, $Nd_2O_3$ and CaAs are present in the sample. Also, we aim to know how many calcium ions have been entered into the 1111-structure of the Nd-Ca0.1 sample and so we need to compute the occupancy number of the calcium and the neodymium ions from XRD analysis. As shown in Table I, the maximum occupation number of the calcium ions in the Nd-Ca0.1 sample is obtained 0.0498 and more than this amount of the calcium cannot be substituted in the neodymium sites. Although we have tried to synthesize the sample with x=0.1 content, the Nd-Ca0.05 phase (5%) forms in this sample because of the calcium solubility restriction. So, we suggest that the solubility of the calcium ions is limited to x≤0.05 in the polycrystalline $Nd_{1-x}Ca_xFeAsO_{0.8}F_{0.2}$ compounds. In a similar result, A. Marcinkova et al. [44] had specified the limit of the calcium solubility for the NdFeAsO sample and it was restricted to x≤0.05. Then, we concentrate to investigate the calcium substitution effects on the Nd-1111 sample for x≤0.05.

The XRD patterns of the synthesized samples that are refined by the MAUD software are presented in Fig. 4. The goodness of fit (S parameter) is characterized by S= $R_{wp}$ /$R_{exp}$, where $R_{wp}$ is the weighted residual error and $R_{exp}$ is the expected error. The S parameters of our samples are listed in Table II, which show the refinements have good quality.

The obtained lattice parameters by employing the MAUD software for the synthesized samples are listed in Table II. It is seen that between the lattice parameters, the decrement of the "c" is further by increasing the calcium content and also, the cell volume decrease. To clarify this issue, we know that there isn't much difference between the ionic radiuses of $Ca^{2+}$ (1.12 Å) and $Nd^{3+}$ (1.11 Å), so we need to calculate other structural parameters. These parameters for our samples have been listed in Tables II and III. It is understood from Table II that the bond lengths of the Nd-(O/F) and the Fe-As decrease with the increase in the calcium doping, which cause a little decrease of the lattice parameter "a". Also, the bond angle of (O/F)-Nd-(O/F) increases with the increase in the calcium content. It can be attributed to the electronegativity difference of the calcium ions in comparison to the neodymium ions. The increase in the (O/F)-Nd-(O/F) angle and the reduction in the bond length of the Nd-(O/F) lead to shrinkage of the Nd-(O/F) layer (are listed in Table III). Also, the change of the (O/F)-Nd-(O/F) angle is effective on the variation of the As-Fe-As angles i.e. "α, and β". Table II indicates that the values of α and β are found to increase and decrease, respectively, by increasing the calcium content. These issues and the decrease in the bond length of the Fe-As are leading to a compression of the Fe-As layer or the "pnictogen height" (see Table III). Figure 5 displays the schematic picture of our samples. Furthermore, we conclude that the contraction of the Fe-As and Nd-(O/F) layers lead to a reduction in the distance between these layers (that are listed in Table III) and consequently decreasing the "c" and the cell volume.

**B. Electrical measurement**

The temperature dependence of $R/R_0$ for our synthesized samples has been displayed in Fig. 6. For the Nd-1111 sample, the electrical resistivity gradually declines by decreasing temperature and then the superconductivity transition occurs at $T_C^{mid} = 53$ K. The Nd-Ca0.01 sample represents the structural transition at 125 K and then the electrical resistivity slowly decreases by cooling. So the superconductivity transition of this sample happens at $T_C^{mid} = 48$ K. As shown in Fig. 6, the structural transition occurs at 145K for the Nd-Ca0.025 sample and then the electrical resistivity decreases by cooling temperature (metallic behavior). Finally, it shows the superconducting transition at $T_C^{mid}$=27 K. T. Nomura et al. [28] asserted the existence of the structural transition suggests that the superconductivity appears in the orthorhombic phase. So, two previous samples experience a superconducting transition in the orthorhombic structure. It is evident from the inset of Fig. 6 that the structural transition doesn't exist for the Nd-Ca0.05 sample. So based on Ref. 28, we expect that it doesn't have the superconducting transition, as it happened exactly in this figure. Consequently, the Nd-Ca0.05 is not a superconductor but it displays a semiconducting behavior. This issue has existed in some other compounds of FeSCs in Refs.45-48.

**C. Dependence of the $T_c$ and the bond angels**

As shown in Table III, the $T_C$ decreases by substitution of the $Ca^{2+}/Nd^{3+}$ ions. In addition, as shown in Fig. 7 (a), the $T_C$ decreases with the deviation of α and β from the regular tetrahedron (α=β=109.47°). Hence, the angular changes in the superconductivity planes influence on the $T_C$. Our results are consistent with Lee's plot [34, 35] and also, H. Usui et al. [32, 33] had studied theoretically the dependence of the $T_C$ and the As-Fe-As bond angle based on the spin-fluctuation theory and the fluctuation exchange (FLEX) method. Accordingly, the eigenvalue $\lambda_E$ of the Eliashberg equation attains unity at the T = Tc, so that it can be used as a qualitative criterion for Tc. Also, the $\lambda_E$ (and so $T_C$) is decreased by moving away of the bond angle from the regular value. Moreover, we have shown in our previous work that the distortions may be independent of the doping type and doping site in the 1111-type of FeSCs[49]. Recently, H. Usui et al. [50] have also examined the variation effect of bond angle on the superconductivity in the 1111-type of FeSCs with isovalent doping, which our experimental data in other previous work confirm it[51]. Finally, the matching of our experimental results and the above-mentioned theory confirms the existence of the spin-fluctuation theory as a pairing mechanism in the FeSCs.

**D. Dependence of the $T_c$ and the pnictogen height**

Based on the theoretical study, K. Kuroki et al. [34-36] have suggested that the Tc decreases by decreasing the pnictogen height ($h_{Pn}$) or the thickness of the Fe-Pn layer that measured from the Fe plane. Also, they showed that the $h_{Pn}$ was indeed the key factor that specified both Tc and the form of the superconducting gap. According to our experimental results that are shown in Fig. 7 (b), we can see that the $T_C$ is decreased by the reduction of $h_{As}$ (i.e. $h_{Pn}$) upon increasing the calcium content in our samples. Therefore our work expresses the experimental aspect of the mentioned theory.

Also, based on two previously mentioned results, the $T_C$ is affected by α and $h_{As}$. So, it is difficult to determine which parameter has a more contribution to the variation of $T_C$. Because these parameters are related to each other by the equation $h_{As}= L_{Fe\text{-}As} \cos(\alpha/2)$, where the $L_{Fe\text{-}As}$ is the bond length of Fe-As.

**E. Dependence of the $T_c$ and the Fe-As bond length**

In Fig. 7 (c), we plot the measured $T_C$ as a function of the Fe-As bond length for the various calcium contents. The $T_C$ decreases with the decrease in the bond length by increasing the calcium doping. Theoretically, H. Usui et al. [32, 33] had studied that the eigenvalue of the Eliashberg equation monotonically increased with the increase of the Fe-As bond length and so the superconductivity enhanced. They obtained that the increment of the density of states originating from the narrowing of the bandwidth was the reason for the superconductivity improvement. Our experimental data is consistent with this result.

**F. Dependence of the $T_c$ and the lattice parameter**

The variation of $T_C$ as a function of the lattice parameter “c” for our synthesized samples is shown in Fig. 7 (d). Accordingly, the $T_C$ reduces with the decrease in the lattice parameter “c” by increasing the calcium content. Theoretically, K. Kuroki et al. [34-36] found that the reduction in the lattice constant “c” suppressed the superconductivity in the 1111-type of FeSCs, which can be attributed to the increase of hopping integrals and associated suppression of the electron correlation based on the spin-fluctuation theory. Hence, our empirical research confirms the above theoretical result.

**G. Dependence of the $T_C$ and microstrain**

As we said before, the lattice parameters and the cell volume decrease with the increase in the calcium content. It leads to the creation of the microstrain in our samples. So, we calculate the microstrain η and crystallite size D through Williamson-Hall equation:

$$\beta \cos \theta / \lambda = K/D + \eta \sin \theta / \lambda \qquad (3)$$

Where K is Scherer's constant, β and θ are full width at half maximum (FWHM) and diffraction angle for each peak, respectively[52]. Williamson-Hall plots for the Nd-1111, Nd-Ca0.01, Nd-Ca0.025, and Nd-Ca0.05 samples are shown in Fig. 8. So, the calculated microstrain and the average crystallite size of our synthesized samples are given in Table III. It is seen that long with decreasing of the $T_C$, the microstrain of our samples increases upon increasing the calcium content. Therefore, it seems that the variation of microstrain can also change the $T_C$ and may cause to suppression of the superconductivity in our samples by increasing the calcium content.

# IV. Conclusions

We have synthesized polycrystalline $Nd_{1-x}Ca_xFeAsO_{0.8}F_{0.2}$ samples through the one-step solid state reaction method. Experimentally, we have investigated structural and electrical properties. Our results are listed as following:

1. Based on the MAUD analysis and synthesizing the several samples, the calcium solubility limit had obtained 0.05.
2. Due to the moving away of the As-Fe-As bond angles from the regular tetrahedron value, the $T_C$ decreased by increasing the calcium content.
3. The decreasing of the pnictogen height, the Fe-As bond length, and the lattice parameter "c" caused to suppression of the superconductivity in our samples with the increase in the calcium content.
4. The microstrain of our samples was increased by substitution of the calcium content, which can be attributed to the lattice constriction.

The agreement of our empirical results and theoretical calculations can enhance the validity of the spin- and the orbital-fluctuation theories as paring mechanisms in the FeSCs**.**

## Acknowledgements

The authors are grateful to Vice Chancellor Research and Technology of Alzahra University for financial supports.

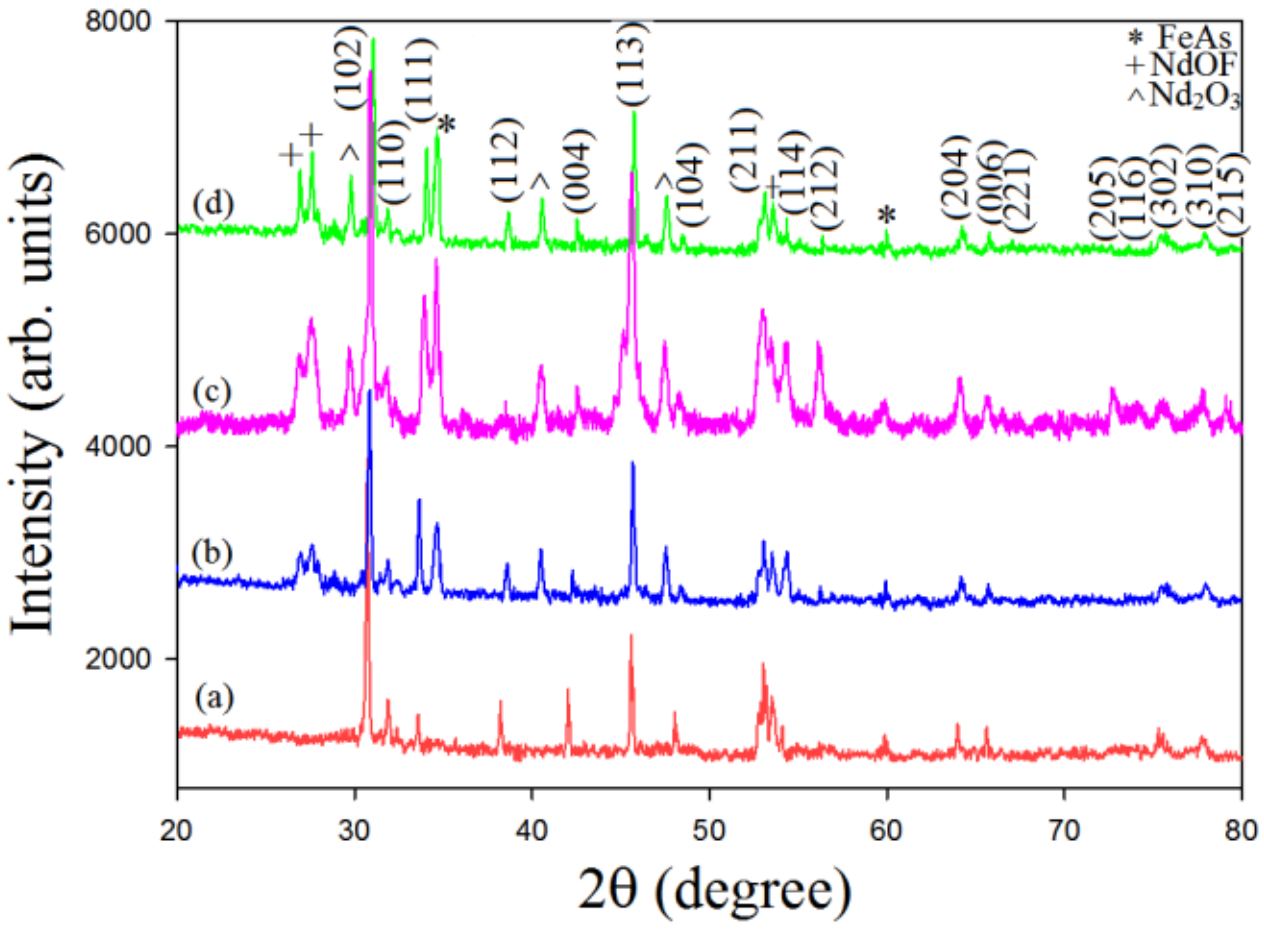

FIG. 1. The XRD patterns of our samples: (a) Nd-1111, (b) Nd-Ca0.01, (c) Nd-Ca0.025 and (d) Nd-Ca0.05

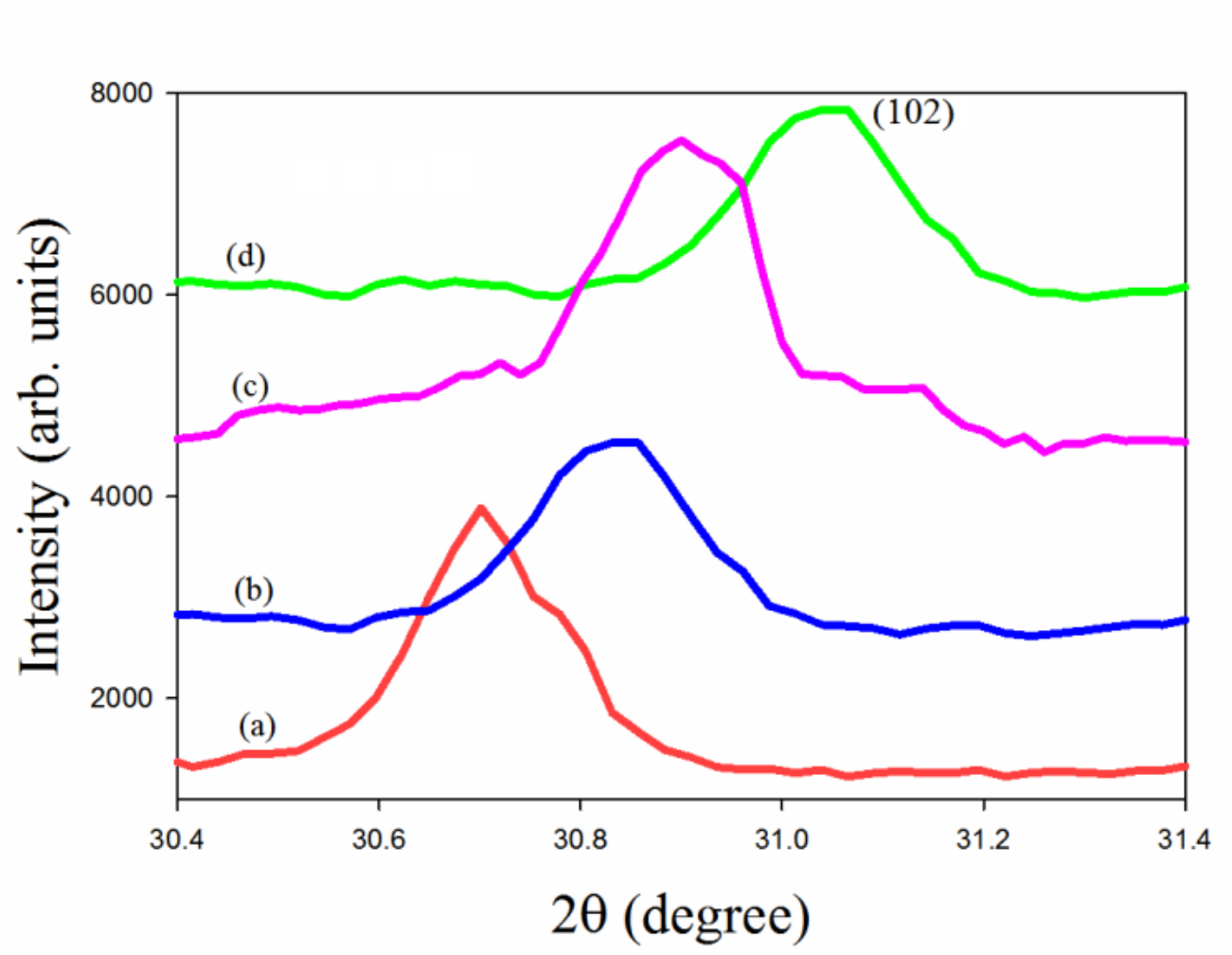

FIG.2. Enlarged view of the (102) peaks for our samples: (a) Nd-1111, (b) Nd-Ca0.01, (c) Nd-Ca0.025 and (d) Nd-Ca0.05

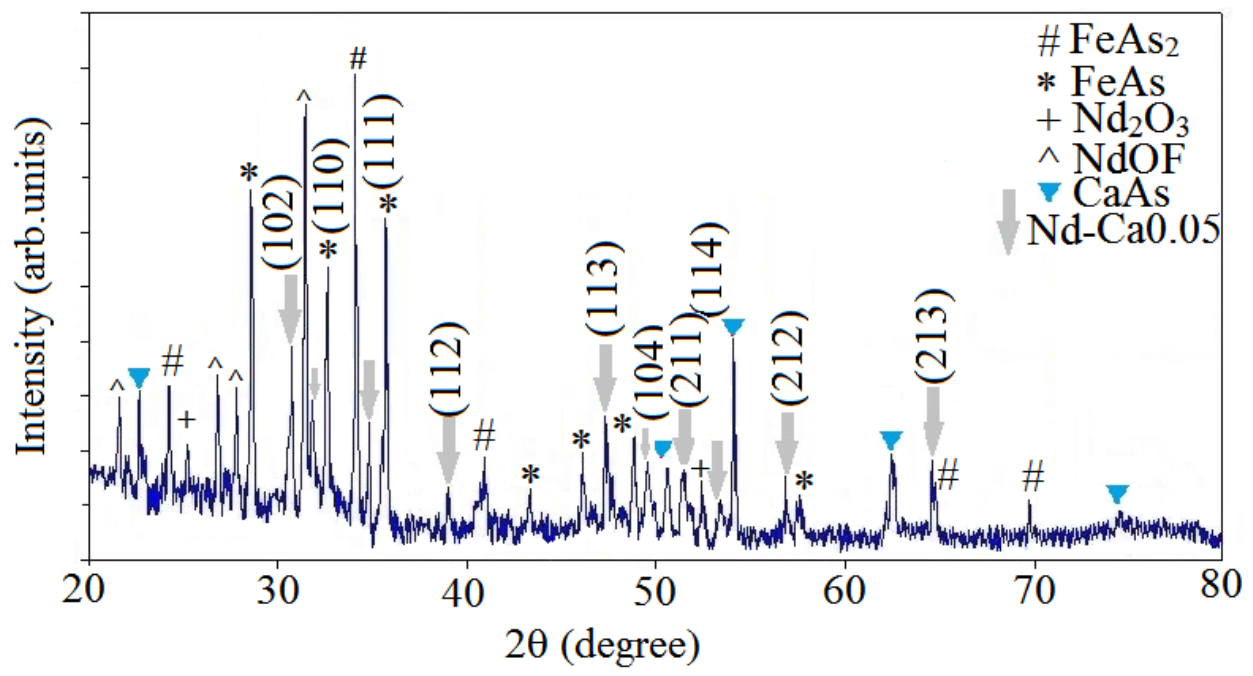

FIG. 3. The XRD pattern of the synthesized Nd-Ca0.1 sample

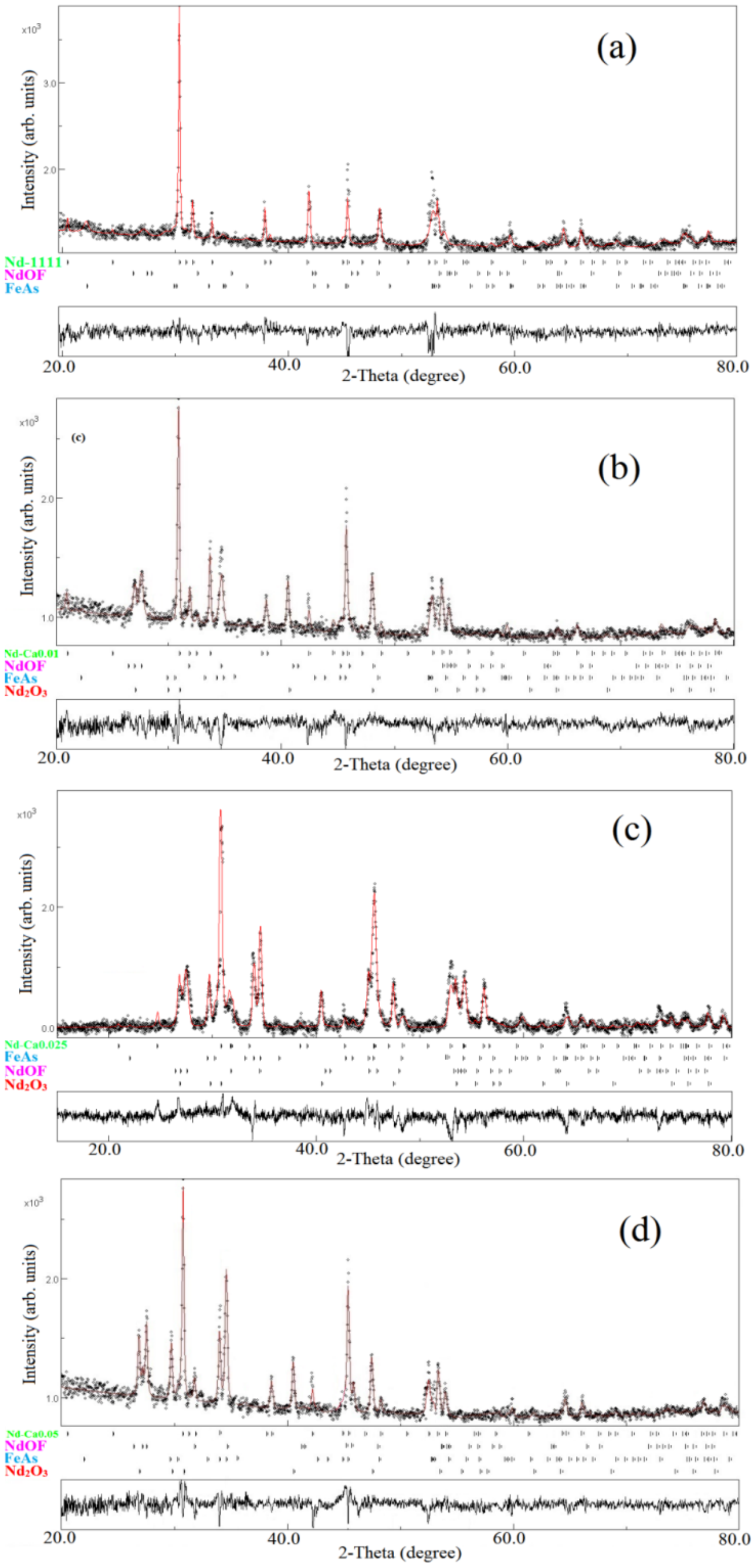

FIG. 4. The refinement of the XRD patterns using MAUD software for (a) the Nd-1111, (b) Nd-Ca0.01, (c) Nd-Ca0.025, and (d) Nd-Ca0.05 samples

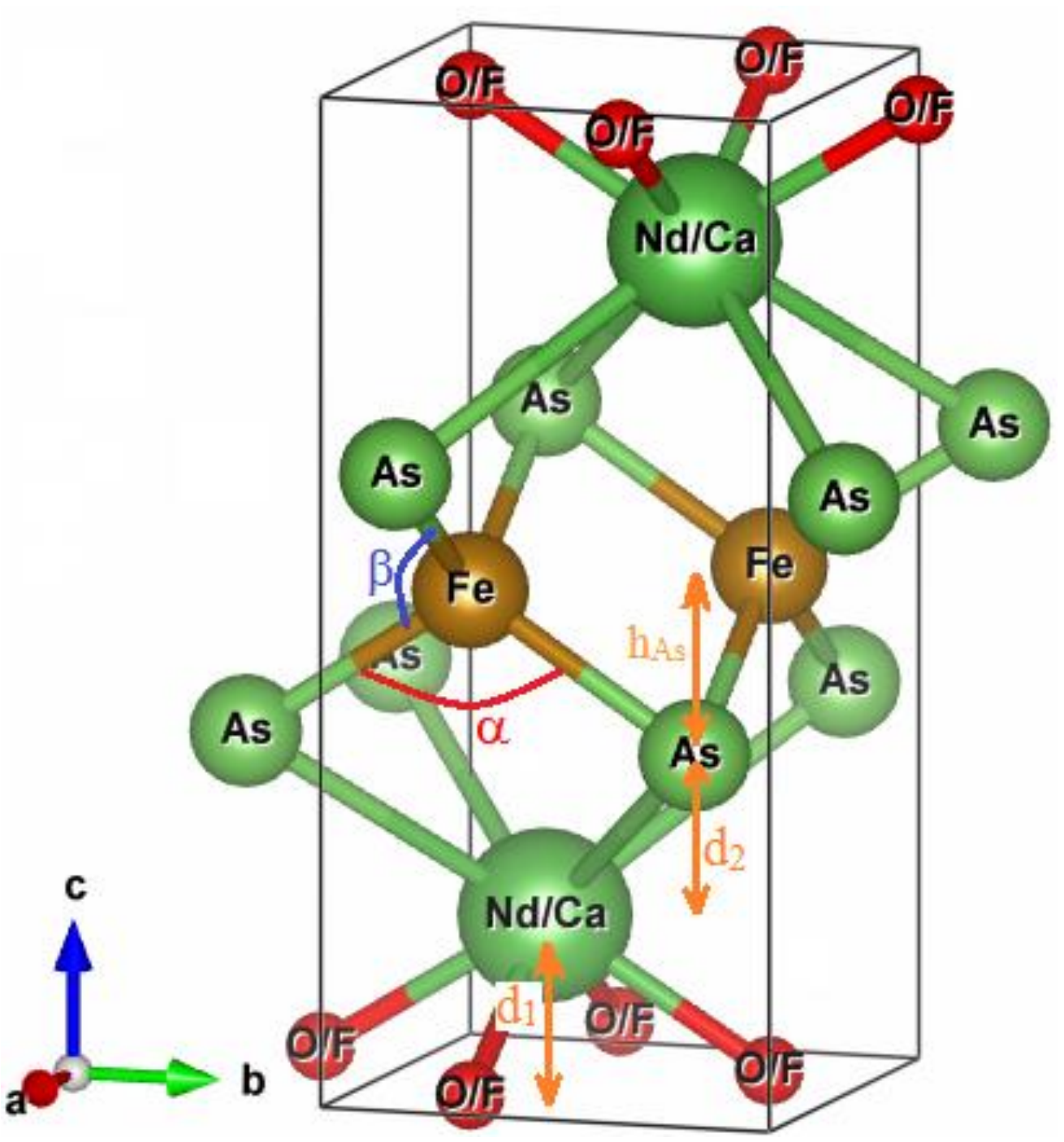

FIG.5. Schematic picture of $Ni_{1-x}Ca_xFeAsO_{0.8}F_{0.2}$ samples

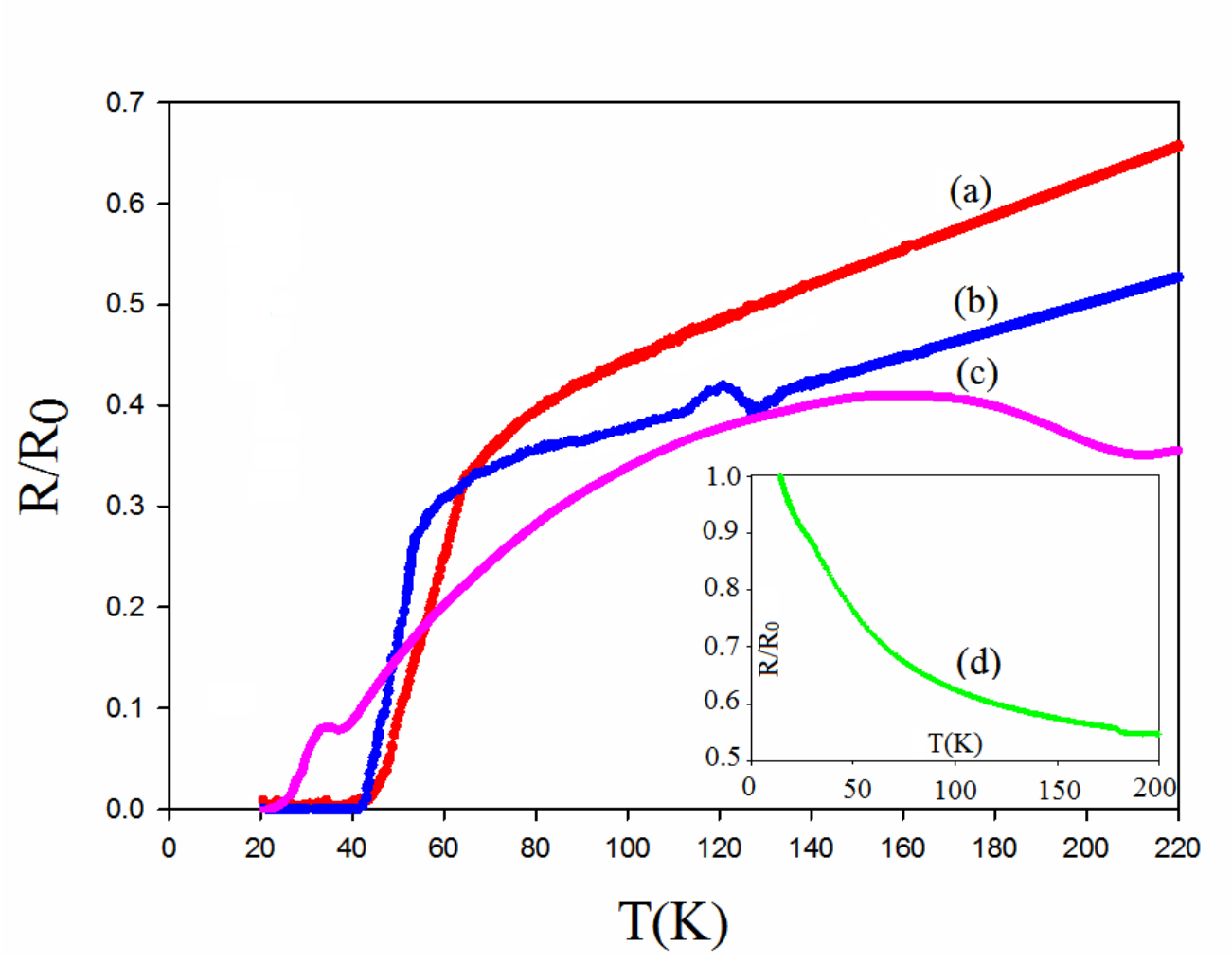

FIG. 6. Temperature dependence of $R/R_0$ for our samples: (a) Nd-1111, (b) Nd-Ca0.01, and (c) Nd-Ca0.025. The inset (d) shows the behavior of the Nd-Ca0.05 sample.

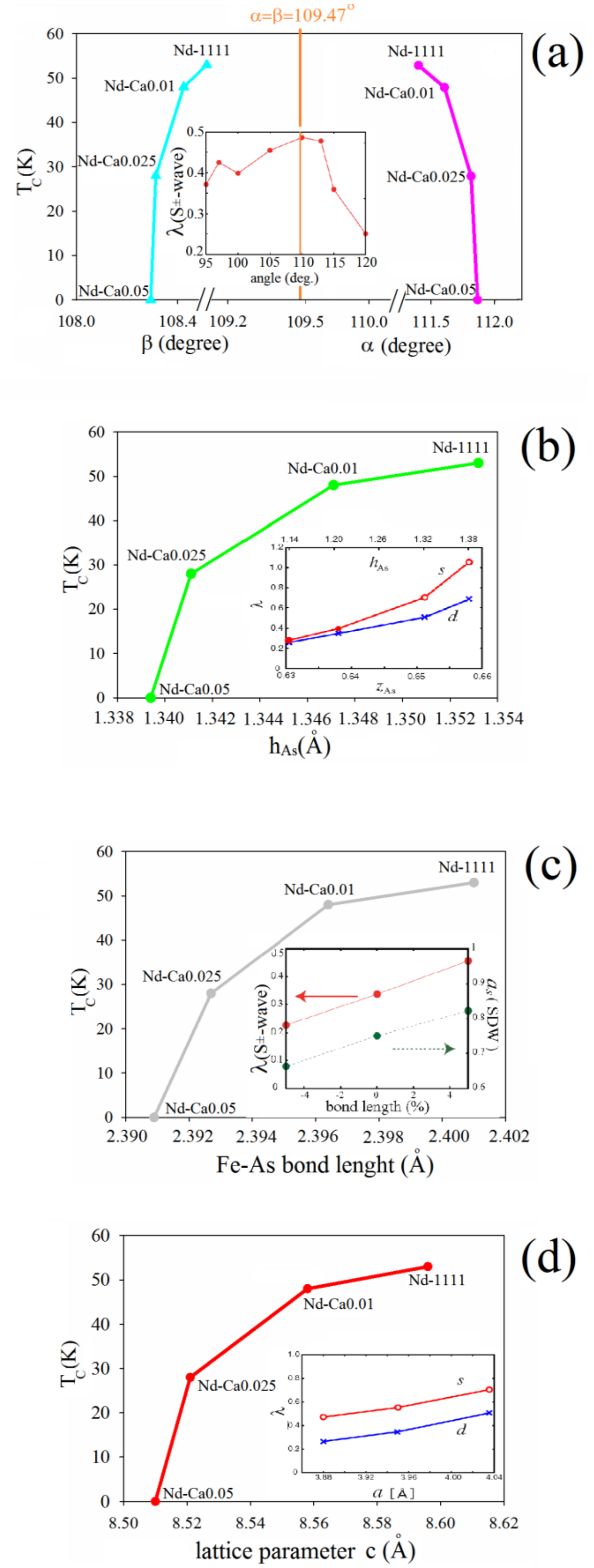

FIG. 7. The variation of $T_C$ as a function of (a) bond angles (Inset of each figure is a theoretical plot of $\lambda_E$ that is extracted from Refs. 33 and 34), (b) $h_{As}$, (c) Fe-As bond length, and (d) c for the various calcium contents.

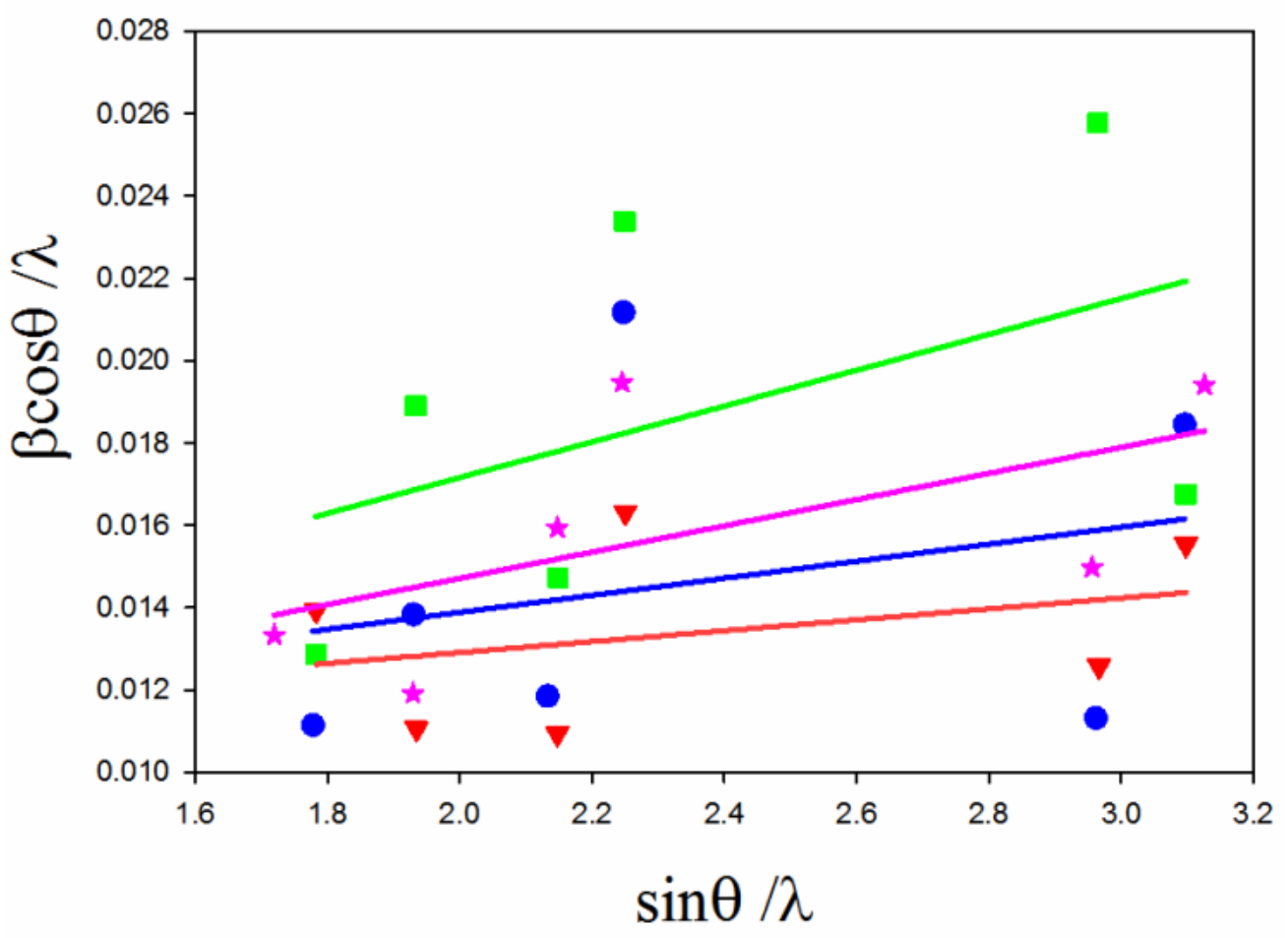

FIG.8. Williamson–Hall plot of Nd-1111(triangle dots), Nd-Ca0.01(circle dots), Nd-Ca0.025 (star dots), and Nd-Ca0.05(square dots) samples

TABLE I. Theoretical- and experimental-refinement values from MAUD analysis for our samples

| **Sample name** | Theoretical-refinement values | | | | | Experimental-refinement values | | | | |
|---|---|---|---|---|---|---|---|---|---|---|
| | **Ions** | **Position** | | | **Occupancy** | **Ions** | **Position** | | | **Occupancy** |
| | | **x** | **y** | **z** | | | **x** | **y** | **z** | |
| **Nd-1111** | $Nd^{3+}$ | 0.25 | 0.25 | 0.1385 | 1 | $Nd^{3+}$ | 0.2499 | 0.2499 | 0.1381 | 0.9998 |
| | $Fe^{2+}$ | 0.75 | 0.25 | 0.5 | 1 | $Fe^{2+}$ | 0.75 | 0.25 | 0.5 | 1 |
| | As | 0.25 | 0.25 | 0.6574 | 1 | As | 0.25 | 0.25 | 0.6574 | 1 |
| | $O^{2-}$ | 0.75 | 0.25 | 0 | 0.8 | $O^{2-}$ | 0.75 | 0.25 | 0 | 0.8 |
| | $F^{-}$ | 0.75 | 0.25 | 0 | 0.2 | $F^{-}$ | 0.75 | 0.25 | 0 | 0.2 |
| **Nd-Ca0.01** | $Ca^{2+}$ | 0.25 | 0.25 | 0.1385 | 0.01 | $Ca^{2+}$ | 0.2498 | 0.2499 | 0.1379 | 0.0105 |
| | $Nd^{3+}$ | 0.25 | 0.25 | 0.1385 | 0.99 | $Nd^{3+}$ | 0.2497 | 0.2499 | 0.1374 | 0.9866 |
| | $Fe^{2+}$ | 0.75 | 0.25 | 0.5 | 1 | $Fe^{2+}$ | 0.75 | 0.25 | 0.5 | 1 |
| | As | 0.25 | 0.25 | 0.6574 | 1 | As | 0.25 | 0.25 | 0.6574 | 1 |
| | $O^{2-}$ | 0.75 | 0.25 | 0 | 0.8 | $O^{2-}$ | 0.75 | 0.25 | 0 | 0.8 |
| | $F^{-}$ | 0.75 | 0.25 | 0 | 0.2 | $F^{-}$ | 0.75 | 0.25 | 0 | 0.2 |
| **Nd-Ca0.025** | $Ca^{2+}$ | 0.25 | 0.25 | 0.1385 | 0.025 | $Ca^{2+}$ | 0.2498 | 0.2498 | 0.1377 | 0.0239 |
| | $Nd^{3+}$ | 0.25 | 0.25 | 0.1385 | 0.975 | $Nd^{3+}$ | 0.2497 | 0.2496 | 0.1372 | 0.9746 |
| | $Fe^{2+}$ | 0.75 | 0.25 | 0.5 | 1 | $Fe^{2+}$ | 0.75 | 0.25 | 0.5 | 1 |
| | As | 0.25 | 0.25 | 0.6574 | 1 | As | 0.25 | 0.25 | 0.6574 | 1 |
| | $O^{2-}$ | 0.75 | 0.25 | 0 | 0.8 | $O^{2-}$ | 0.75 | 0.25 | 0 | 0.8 |
| | $F^{-}$ | 0.75 | 0.25 | 0 | 0.2 | $F^{-}$ | 0.75 | 0.25 | 0 | 0.2 |
| **Nd-Ca0.05** | $Ca^{2+}$ | 0.25 | 0.25 | 0.1385 | 0.05 | $Ca^{2+}$ | 0.2497 | 0.2499 | 0.1376 | 0.0489 |
| | $Nd^{3+}$ | 0.25 | 0.25 | 0.1385 | 0.95 | $Nd^{3+}$ | 0.2496 | 0.2496 | 0.1370 | 0.9506 |
| | $Fe^{2+}$ | 0.75 | 0.25 | 0.5 | 1 | $Fe^{2+}$ | 0.75 | 0.25 | 0.5 | 1 |
| | As | 0.25 | 0.25 | 0.6574 | 1 | As | 0.25 | 0.25 | 0.6574 | 1 |
| | $O^{2-}$ | 0.75 | 0.25 | 0 | 0.8 | $O^{2-}$ | 0.75 | 0.25 | 0 | 0.8 |
| | $F^{-}$ | 0.75 | 0.25 | 0 | 0.2 | $F^{-}$ | 0.75 | 0.25 | 0 | 0.2 |
| **Nd-Ca0.1[a]** | $Ca^{2+}$ | 0.25 | 0.25 | 0.1385 | 0.05 | $Ca^{2+}$ | 0.2496 | 0.2499 | 0.1379 | 0.0498 |
| | $Nd^{3+}$ | 0.25 | 0.25 | 0.1385 | 0.95 | $Nd^{3+}$ | 0.2495 | 0.2496 | 0.1371 | 0.9497 |
| | $Fe^{2+}$ | 0.75 | 0.25 | 0.5 | 1 | $Fe^{2+}$ | 0.75 | 0.25 | 0.5 | 1 |
| | As | 0.25 | 0.25 | 0.6574 | 1 | As | 0.25 | 0.25 | 0.6574 | 1 |
| | $O^{2-}$ | 0.75 | 0.25 | 0 | 0.8 | $O^{2-}$ | 0.75 | 0.25 | 0 | 0.8 |
| | $F^{-}$ | 0.75 | 0.25 | 0 | 0.2 | $F^{-}$ | 0.75 | 0.25 | 0 | 0.2 |

[a] see text for detail.

TABLE II. Different structural parameters for the various calcium contents from MAUD analysis

| Ca content (x) | Vol. %. of phases | | | | S | Lattice parameters | | Cell volume (Å$^3$) | Bond length (Å) | | Bond angle (°) | | |
|---|---|---|---|---|---|---|---|---|---|---|---|---|---|
| | pure | FeAs | NdOF | $Nd_2O_3$ | | a (Å) | c (Å) | | Fe-As | Nd-(O/F) | As-Fe-As | | (O/F)-Nd-(O/F) |
| | | | | | | | | | | | α | β | |
| **x=0** | 81.5 | 10.7 | 7.8 | 0 | 1.719 | 3.967 | 8.596 | 135.276 | 2.4010 | 2.3134 | 111.402 | 108.515 | 116.950 |
| **x=0.01** | 77.0 | 9.8 | 8.7 | 4.5 | 1.867 | 3.964 | 8.558 | 134.474 | 2.3964 | 2.3094 | 111.607 | 108.424 | 117.138 |
| **x=0.025** | 74.2 | 11.4 | 8.5 | 5.9 | 2.185 | 3.963 | 8.521 | 133.825 | 2.3927 | 2.3063 | 111.817 | 108.314 | 117.345 |
| **x=0.05** | 73.8 | 11.7 | 7.7 | 6.7 | 1.365 | 3.961 | 8.510 | 133.517 | 2.3909 | 2.3047 | 111.867 | 108.294 | 117.386 |

TABLE III. Some structural and superconducting parameters for the various calcium contents (see Fig.5 for detail)

| Ca content (x) | Thickness of Fe-As layer or pnictogen height $h_{As}$ (Å) | Thickness of Nd-(O/F) layer $d_1$ (Å) | Distance between Fe-As and Nd-(O/F) layers $d_2$ ( Å) | Williamson-Hall equation | Crystallite size (nm) | Microstrain η (%) | $T_C$ (K) | $T_S$ (K) |
|---|---|---|---|---|---|---|---|---|
| **x=0** | 1.3532 | 1.2096 | 1.7358 | y= 0.0013x+ 0.0103 | 97.09±0.04 | 0.13±0.02 | 53 | - |
| **x=0.01** | 1.3471 | 1.2043 | 1.7276 | y=0.0021x+0.0098 | 102.41±0.06 | 0.21±0.04 | 48 | 125 |
| **x=0.025** | 1.3411 | 1.1991 | 1.7203 | y=0.0032x+0.0072 | 120.48±0.05 | 0.32±0.02 | 27 | 145 |
| **x=0.05** | 1.3394) | 1.1975 | 1.7181 | y= 0.0044x+ 0.0085 | 117.65±0.09 | 0.44±0.09 | - | - |